\documentstyle[eqsecnum,aps,prd,epsfig]{revtex}
\date{\today}
\tighten
\begin{document}
\tighten
%%%%%%%%%%%%%%% macros %%%%%%%%%%%%%%%%%%%%%%%%%
%
\let\ov=\over
\let\lbar=\l
\let\l=\left
\let\r=\right
\def \der#1#2{{\partial{#1}\over\partial{#2}}}
\def \dder#1#2{{\partial^2{#1}\over\partial{#2}^2}}
\def\N{{I\!\!N}}
\def\be{\begin{equation}}
\def\ee{\end{equation}}
\def\beu{\begin{displaymath}}
\def\eeu{\end{displaymath}}
\def\bea{\begin{eqnarray}}
\def\eea{\end{eqnarray}}
\def\beau{\begin{eqnarray*}}
\def\eeau{\end{eqnarray*}}
\def\ms{\langle S \rangle}
\def\n2{\langle N^2 \rangle}
\def\sn2{\sqrt{\langle N^2 \rangle}}
\def\laq{\raise 0.4ex\hbox{$<$}\kern -0.8em\lower 0.62
ex\hbox{$\sim$}}
\def\gaq{\raise 0.4ex\hbox{$>$}\kern -0.7em\lower 0.62
ex\hbox{$\sim$}}
% 
%%%%%%%%%%%%%%%%%%%%%%%%%%%%%%%%%%%%%%%%%%%%%%%%%%%%%%%%
%
\title{Detecting relic gravitational radiation  from string cosmology
with LIGO}
\author{Bruce Allen}
\address{Department of Physics\\
University of Wisconsin - Milwaukee\\
PO Box 413\\
Milwaukee, WI 53211, USA\\
email: ballen@dirac.phys.uwm.edu}

\author{Ram Brustein}
\address{Department of Physics\\
Ben-Gurion University\\
Beer-Sheva 84105, Israel\\
email: ramyb@bgumail.bgu.ac.il}

\maketitle
\begin{abstract}
A characteristic spectrum of relic gravitational radiation is produced
by a period of ``stringy inflation" in the early universe.  This
spectrum is unusual, because the energy-density rises rapidly with
frequency.  We show that correlation experiments with the two
gravitational wave detectors being built for the Laser Interferometric
Gravitational Observatory (LIGO) could detect this relic radiation, for
certain ranges of the parameters that characterize the underlying
string cosmology model.
\end{abstract}

\vskip -0.1in
\centerline{Preprint Numbers: WISC-MILW-96-TH-34,BGU-PH-96/09}
\section{INTRODUCTION}
Because the gravitational interaction is so weak, a stochastic
background of gravitational radiation (the graviton background)
decouples from the matter in the universe at very early times.  For
this reason, the stochastic background of gravitational radiation,
which is in principle observable at the present time, carries with it a
picture of the state of the universe at very early times, when energy
densities and temperatures were very large.

The most interesting features of string theory  are associated with its
behavior at very high energies, near the Planck scale.  Such high
energies are unobtainable in present-day laboratories, and are unlikely
to be reached for quite some time. They were, however, available during
the very early history of the universe, so string theory can be probed
by the predictions which it makes about that epoch.

Recent work has shown how the early universe might behave, if
superstring theories are a correct description of nature \cite{1,2}.
One of the robust predictions of this ``string cosmology" is that our
present-day universe would contain a stochastic background of
gravitational radiation \cite{bggv,gg}, with a spectrum which is quite
different than that predicted by many other early-universe cosmological
models \cite{grishchuk,turner,myreview}.  In particular, the spectrum
of gravitational waves predicted by string cosmology has rising
amplitude with increasing frequency.  This means that the radiation
might have large enough amplitude to be observable by ground-based
gravity-wave detectors, which operate at frequencies above $\approx 10
\> \rm Hz$.  This also allows the spectrum to be consistent with
observational bounds arising at $10^{-18} \> \rm Hz$ from observations
of the Cosmic Background Radiation and at $10^{-8} \> \rm Hz$ from
observation of millisecond pulsar timing residuals.
 
In this short paper, we examine the spectrum of radiation produced by
string cosmology, and determine the region of parameter space for which
this radiation would be observable by the (initial and advanced
versions of the) LIGO detectors \cite{science92}.

\section{Stochastic background in string cosmology}
\label{s:first}
In models of string cosmology \cite{bggv}, the universe passes through
two early inflationary stages.  The first of these is called the
``dilaton-driven" period and the second is the ``string" phase.  Each
of these stages produces stochastic gravitational radiation; the
contribution of the dilaton-driven phase is currently better understood
than that of the string phase.

In order to describe the background of gravitational radiation, it is
conventional to use a spectral function $\Omega_{\rm GW}(f)$ which is
determined by the energy density of the stochastic gravitational
waves.  This function of frequency is defined by
\be \label{e:defomegagw}
\Omega_{\rm GW}(f) = {1 \over \rho_{\rm critical}} {d \rho_{\rm
GW} \over d \ln f}.
\ee
Here $ d \rho_{\rm GW}$ is the (present-day) energy density in
stochastic gravitational waves in the frequency range $ d \ln f$, and
$\rho_{\rm critical}$ is the critical energy-density required to just
close the universe.  This is given by
\be \label{e:crit}
\rho_{\rm critical} = { 3 c^2 H_0^2 \over 8 \pi G} \approx 1.6 \times
10^{-8} {\rm h}_{100}^2 \rm \> ergs/cm^3,
\ee
where the Hubble expansion rate $H_0$ is the rate at which our universe
is currently expanding,
\be \label{e:hubble}
H_0 = {\rm h}_{100} \> 100 \> {\rm  Km \over sec-Mpc} = 
3.2 \times 10^{-18}{\rm h}_{100} {\rm 1 \over sec} .
\ee
Because $H_0$ is not known exactly, it is defined in terms of a
dimensionless parameter ${\rm h}_{100}$ which is believed to lie in the
range $1/2 < {\rm h}_{100} < 1$.

\begin{figure}
\begin{center}
\epsfig{file=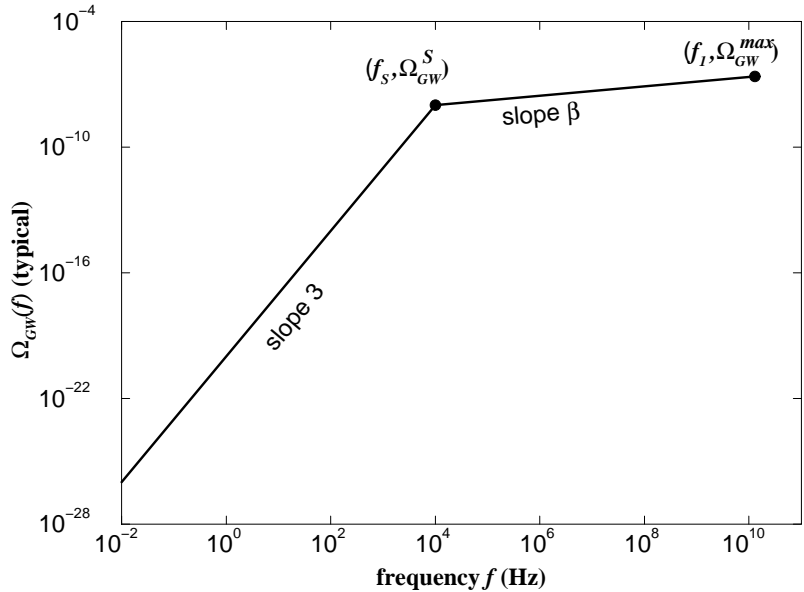, width=12cm, bbllx=0pt, bblly=0pt,
bburx=560pt, bbury=460pt}
\end{center}
\caption{
\label{f:spec}
A  string cosmology gravitational wave  power spectrum.  The part of
the spectrum with slope 3 below frequency ${f_{\rm S}}$ is produced
during the dilaton-driven phase, and the part of the spectrum above
that frequency is produced during the string phase.}
\end{figure}

The spectrum of gravitational radiation produced in the dilaton-driven
and string phase was discussed in \cite{bggv}.  In the simplest model,
which we will use in this paper, it depends upon four parameters.  The
first pair of these are the  frequency ${f_{\rm S}}$ and the fractional
energy density $\Omega^{\rm S}_{\rm GW} $  produced at the end of the
dilaton-driven phase.  The second pair of parameters are the maximal
frequency $f_1$  above which gravitational radiation is not produced
and the maximum fractional energy density $ \Omega^{\rm max}_{GW} $,
which occurs at that frequency.  This is illustrated in
Fig.~\ref{f:spec}.
\noindent

An approximate form for the spectrum is \cite{rb}
\be \label{e:approx}
\Omega_{\rm GW}(f)=
\cases{
{\Omega^{\rm S}_{\rm GW} (f/{f_{\rm S}})^3}  &  { $f<{f_{\rm S}}$} \cr
 & \cr
{\Omega^{\rm S}_{\rm GW} (f/{f_{\rm S}})^{\beta} } &   ${f_{\rm
S}}<f<f_1$ \cr
 & \cr
{0} &   $f_1<f$}
\ee
where
\beu
\beta=\frac{\log\left[\Omega_{\rm GW}^{\rm max}/
\Omega^{\rm S}_{\rm GW}\right]}{\log\left[f_1/{f_{\rm S}}\right]}
\eeu
is the logarithmic slope of the spectrum produced in the string phase.

If we assume that there is no late entropy production and make
reasonable choices for the number of effective degrees of freedom, then
two of the four parameters may be determined in terms of the Hubble
parameter $H_{\rm r}$ at the onset of radiation domination immediately
following the string phase of expansion \cite{bgv},
\be
f_1= 1.3 \times 10^{10} \> {\rm Hz} \left( { H_{\rm r} \over 5 
\times 10^{17} \> {\rm GeV}} \right)^{1/2}
\ee
and
\be
\Omega_{\rm GW}^{\rm max} = 1 \times 10^{-7} {\rm h}_{100}^{-2} 
\left( { H_{\rm r} \over 5 \times 10^{17} \> {\rm GeV}} \right)^2.
\ee
More complicated models and spectra were discussed in
\cite{mg,maggiore,occ}.

The ratios $ \left( \Omega^{\rm S}_{\rm GW}/\Omega_{\rm GW}^{\rm max}
\right)$ and $ \left( f_{\rm S}/f_1\right)$ are determined by the basic
physical parameters of string cosmology models, the values of the
Hubble parameter and the string coupling parameter at the end of the
dilaton-driven phase and the onset of the string phase
\cite{bggv,rb,v96}.

\section{Detecting a stochastic background}
\label{s:second}
A number of authors \cite{mich,chris,flan,myreview} have shown how one
can use a network of two or more gravitational wave antennae to detect
a stochastic background of gravitational radiation.  The basic idea is
to correlate the signals from separated detectors, and to search for a
correlated strain produced by the gravitational-wave background, which
is buried in the intrinsic instrumental noise.  It has been shown by
these authors that after correlating signals for time $T$ (we take
$T=10^7 \> {\rm sec} = 3 \> {\rm  months}$) the ratio of ``Signal" to
``Noise" (squared) is given by an integral over frequency $f$:
\be \label{e:sovern}
\left( {S \over N} \right)^2 =
{9 H_0^4 \over 50 \pi^4} T \int_0^\infty df \>
{\gamma^2 (f) \Omega_{\rm GW}^2(f) \over f^6 P_1(f) P_2(f)}.
\ee

\begin{figure}
\begin{center}
\epsfig{file=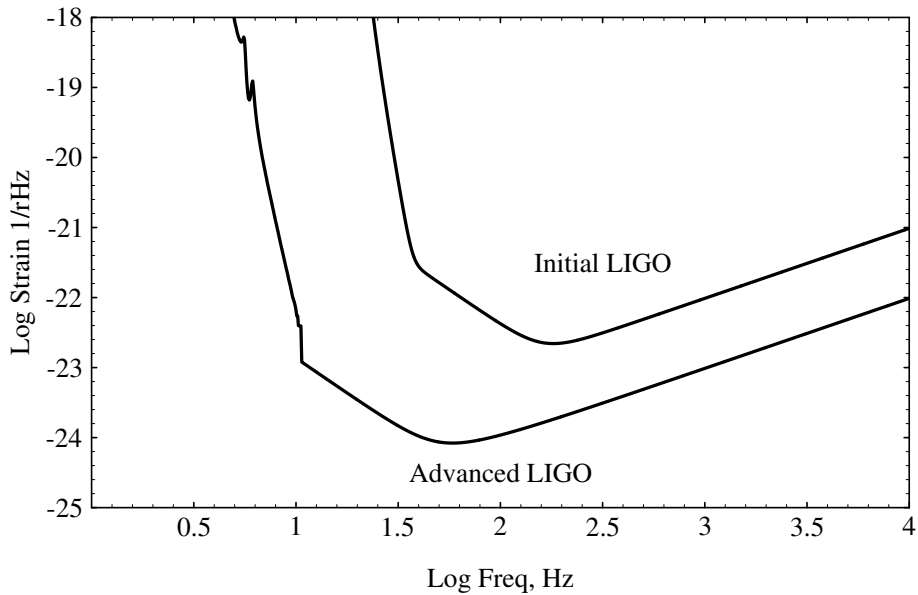, width=12cm, bbllx=74pt, bblly=248pt,
bburx=537pt, bbury=565pt}
\end{center}
\caption{
\label{f:noise}
\noindent
The predicted noise power spectra of the initial and advanced LIGO
detectors.   The horizontal axis is $\rm log_{10}$ of frequency $f$, in
Hz.  The vertical axis shows $\rm log_{10}  (P(f)/sec)^{1/2}$, or
strain per root Hz.  These noise power spectra are the published design
goals.  The bumps appearing in the low-frequency part of the advanced
LIGO noise curve are obtained by folding measured seismic noise data
with the predicted transfer function of the seismic isolation (stack)
system.}
\end{figure}

\begin{figure}
\begin{center}
\epsfig{file=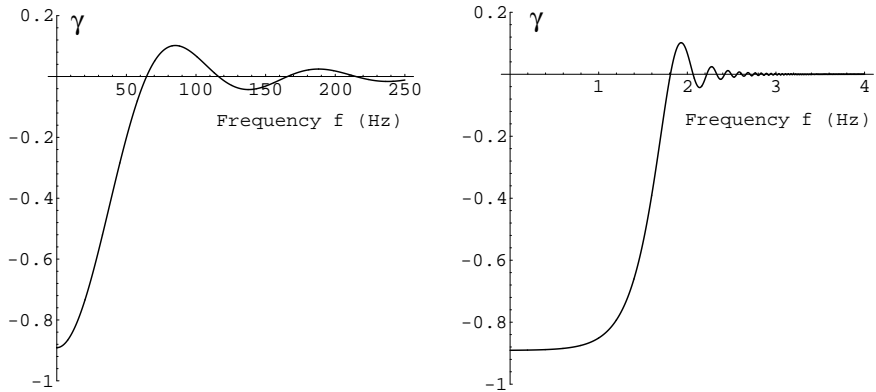,width=12cm,bbllx=2.8cm,bblly=10.3cm,
bburx=19.2cm,bbury=17.5cm}
\end{center}
\caption{
\label{f:overlap}
The overlap reduction function $\gamma(f)$ for the two LIGO detector
sites.  (The horizontal axis of the left-hand graph is linear, while
that of the right-hand graph is $\rm log_{10}$.)  The overlap reduction
function shows how the correlation of the detector pair to an
unpolarized stochastic background falls off with frequency.  The
overlap reduction function has its first zero at 64 Hz, and falls off
rapidly at higher frequencies.}
\end{figure}

In order to detect a stochastic background with $90\%$ confidence the
ratio $S \over N$ needs to be at least $1.65$.  In this equation,
several different functions appear, which we now define.
The instrument noise in the detectors is described by the one-sided
noise power spectral densities $P_i(f)$.  The LIGO project is building
two identical detectors, one in Hanford Washington and one in
Livingston Louisiana, which we will refer to as the ``initial"
detectors.  After several years of operation, these detectors will be
upgraded to so-called ``advanced" detectors.   Since the two detectors
are identical in design, $P_1(f)=P_2(f)$.  The design goals for the
detectors specify these functions \cite{science92}.  They are shown in
Fig.~\ref{f:noise}. The next quantity which appears is the overlap reduction function
$\gamma(f)$.  This function is determined by the relative locations and
orientations of the arms of the two detectors, and is identical for
both the initial and advanced LIGO detectors.  For the pair of LIGO
detectors
\be
\gamma(f) = -0.124842 \> j_{0}(x) - 2.90014 \> {{j_{1}(x)}\over x} + 
    3.00837 \> {{j_{2}(x)}\over {{x^2}}},
\ee
where the $j_i$ are spherical Bessel functions.  The dimensionless
frequency variable is $x=2 \pi f \tau$ with $\tau=10.00 \> \rm msec$
being the light-travel-time between the two LIGO detector sites.  This
function is shown in Fig.~\ref{f:overlap}.

\noindent
Equation (\ref{e:sovern}) allows us to assess the detectability (using
initial or advanced LIGO) of any particular stochastic background
$\Omega_{\rm GW}(f)$.

\section{Detecting a string cosmology stochastic background}
\label{s:third}
\begin{figure}
\begin{center}
\epsfig{file=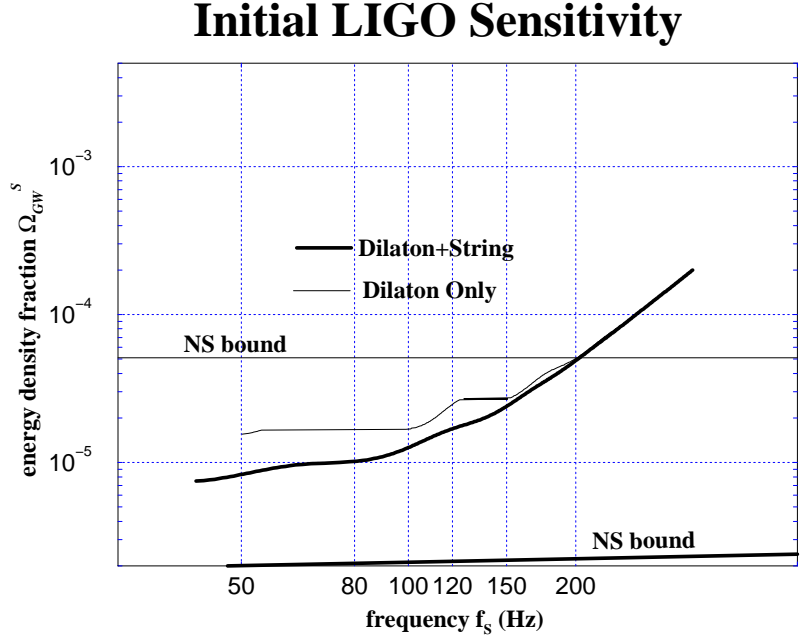, width=12cm, bbllx=0pt, bblly=0pt,
bburx=560pt, bbury=460pt}
\end{center}
\caption{
\label{f:init}
Shown is the region in parameter space for which the gravitational wave
stochastic background is observable by the initial LIGO detector.  The
region above and to the left of the curves is where the signal to noise
ratio exceeds 1.65 and the background is observable with 90\%
confidence. 
The two curves are indistinguishable above $f_{\rm S} = 200 \> \rm  Hz$.
The two lines labeled ``NS bound" are nucleosynthesis
bounds.  The allowed region is below each line. Note that the
Dilaton+String line lies below the 90\% confidence curve.}
\end{figure}

\begin{figure}
\begin{center}
\epsfig{file=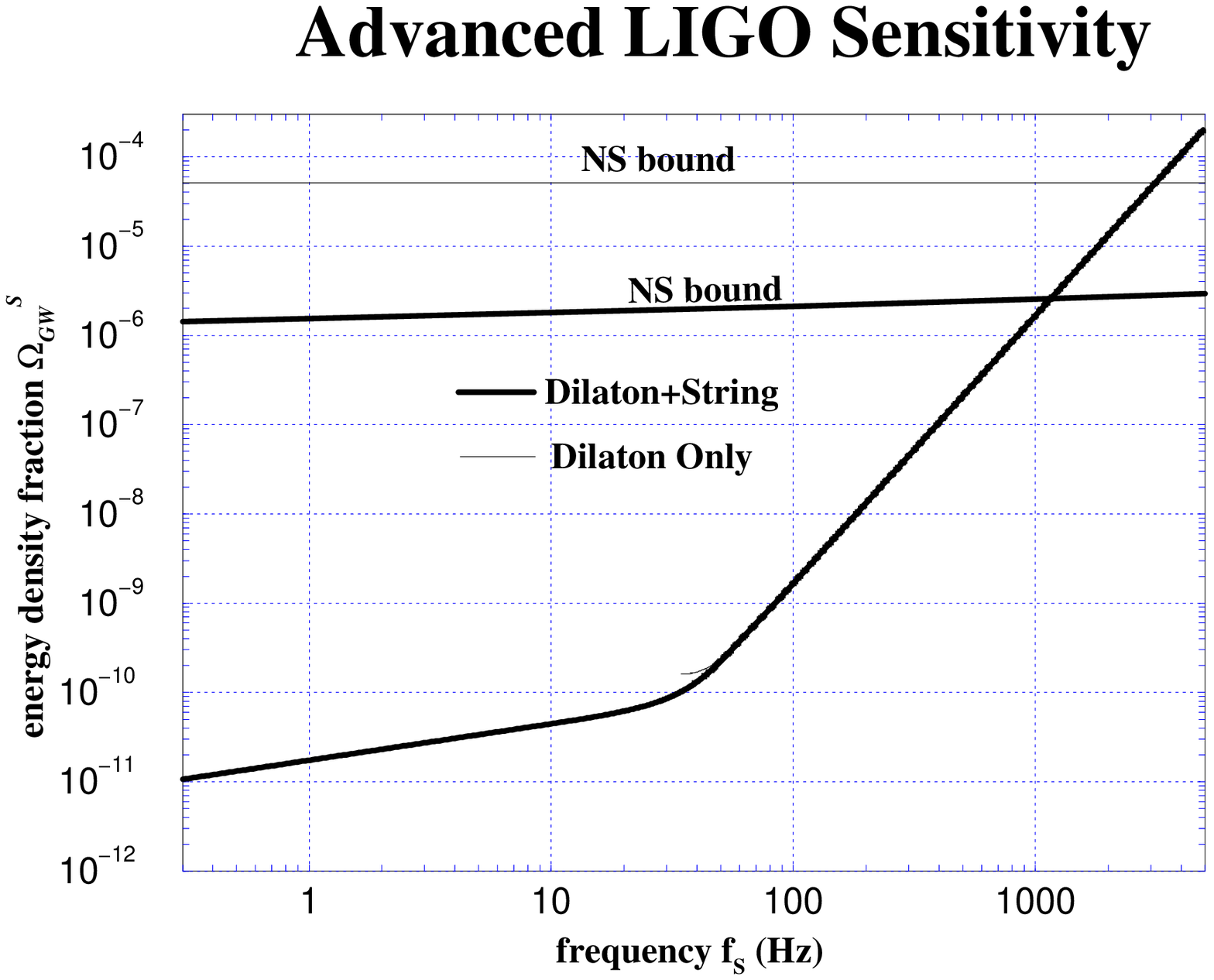, width=12cm, bbllx=0pt, bblly=0pt,
bburx=560pt, bbury=460pt}
\end{center}
\caption{
\label{f:adva}
This graph is identical to the previous one, but for the advanced LIGO
detector.  The observable region of parameter space is above and to the
left of the curves. 
The two curves are indistinguishable above $f_{\rm S} = 50 \> \rm Hz$.}
\end{figure}

Making use of the prediction from string cosmology, we may use equation
(\ref{e:sovern}) to assess the detectability of this stochastic
background.  For any given set of parameters we may numerically
evaluate the signal to noise ratio $S/N$; if this value is greater than
$1.65$ then with at least 90\% confidence, the background can be
detected by LIGO.  The regions of detectability in parameter space are
shown in Fig.~\ref{f:init} for the initial LIGO detectors, and in
Fig.~\ref{f:adva} for the advanced LIGO detectors.  For these figures
we have assumed ${\rm h}_{100}=0.65$ and $H_{\rm r}=5 \times 10^{17} \>
{\rm GeV}$. The observable regions for different values of these
parameters can be obtained by simple scaling of the presented results.

At the moment, the most restrictive observational constraint on the
spectral parameters comes  from  the standard model of big-bang
nucleosynthesis (NS) \cite{ns1}.  This restricts the total energy
density in gravitons to less than that of approximately one massless
degree of freedom in thermal equilibrium. This bound implies that
\be
\int \Omega_{\rm GW}(f) d \ln f  = \Omega_{\rm GW}^{\rm S} \left[
\frac{1}{3}+ \frac{1}{\beta}\left( \left(f_1/f_{\rm S}
\right)^\beta-1\right)\right] < 0.7 \times 10^{-5} {\rm h}^{-2}_{100}.
\label{nucleo}
\ee
where we have assumed an allowed $N_\nu=4$ at NS, and have substituted
in the spectrum (\ref{e:approx}).  This bound is shown on
Figs.~\ref{f:init},\ref{f:adva}.  We also show the weaker ``Dilaton Only"
bound, assuming NO stochastic background is produced during the (more
poorly-understood) string phase of expansion:
\be
\Omega_{\rm GW}^{\rm S} < 2.1 \times 10^{-5} {\rm h}^{-2}_{100}.
\ee
This is obtained by setting $f_1=f_{\rm S}$ in the previous equation, i.e. assuming
that $\Omega_{\rm GW}$ vanishes for ${f_{\rm S}} <f < f_1$.  We note
that if the ``Dilaton + String" spectrum is correct, then the NS bounds
rule out any hopes of observation by initial LIGO.  On the other hand,
in the ``Dilaton Only" case, a detectable background is not ruled out
by NS bounds; it would be observable if the spectral peak falls into
the detection bandpass between 50 and 200 Hz (figure 7 of
\cite{myreview}).  

Because the function  $\left(\frac{\gamma(f)}{P(f)}\right)^2$ decays
rapidly at high and low frequencies, the asymptotic behavior of the
90\% confidence contours for the advanced LIGO detector in the high and
low $f_{\rm S}$ regions can be estimated analytically  as follows
\be
(\Omega_{\rm GW}^{\rm S})_{90\%}= 7.4 \times 10^{-10} {\rm
h}_{100}^{-2} (f_{\rm S}/100 \> {\rm Hz})^{3}, \ \ \ \ f_{\rm S}\gaq
100 \> {\rm Hz}
\ee
\be
(\Omega_{\rm GW}^{\rm S})_{90\%}=  2.6 \times 10^{-11}\ {\rm
h}_{100}^{-2}\ (f_{\rm S}/20 \> {\rm Hz})^{0.4}, \ \ \ \ f_{\rm S}\laq
20 \> {\rm Hz}
\ee
Similar estimates may be obtained for the initial LIGO detector.

\section{Conclusion}
In this short paper, we have shown how data from the the initial and
advanced versions of LIGO will be able to constrain string cosmology
models of the early universe.  In principle, this might also constrain
the fundamental parameters of superstring theory.  The initial LIGO is
sensitive only to a narrow region of parameter space and is only
marginally above the required sensitivity, while the advanced LIGO
detector has far better detection possibilities.  The simultaneous
operation of other types of gravitational wave detectors which operate
at higher frequencies, such as bar and resonant designs, ought to
provide additional increase in sensitivity and therefore further
constrain the parameter space.
 
\acknowledgements
This work has been partially supported by the National Science
Foundation grant PHY95-07740 and by the Israel Science Foundation
administered by the Israel Academy of Sciences and Humanities.

%%%   References
%

\end{document}